\documentclass[journal]{IEEEtran}

\usepackage[svgnames]{xcolor}
\usepackage{graphicx}
\usepackage[font=small, skip=5pt, belowskip=0pt]{caption}
\usepackage[font=footnotesize]{subcaption}
\usepackage{multirow}
\usepackage{array}
\usepackage{booktabs}
\usepackage{pdflscape}
\usepackage[ruled,vlined,linesnumbered]{algorithm2e}
\usepackage[T1]{fontenc}
\usepackage[utf8]{inputenc}
\usepackage{cite}
\usepackage{url}
\usepackage{comment}
\usepackage{ragged2e}
\usepackage{mleftright,xparse}
\usepackage{amssymb,amsmath,amsfonts,amsthm,bm,mathtools,cuted,bbold}
\usepackage{siunitx}

\usepackage[outline]{contour}
\usepackage{tikz}
\usetikzlibrary{math, shapes, calc, patterns, fadings}
\usepackage{pgfplots}
\usepgfplotslibrary{groupplots}
\pgfplotsset{
compat=1.13,
legend style={font=\footnotesize, fill opacity=0.7,  draw opacity=1, text opacity=1, draw=white!15!black, legend cell align=left, align=left}, 
width=6cm, 
height=6cm,
yminorticks=false,
xminorticks=false,
label style={font=\small},
title style={font=\small},
tick style={color=black},
tick align=outside,
tick pos=left,
tick label style={font=\footnotesize},
grid style={line width=.1pt, draw=gray!20},
major grid style={line width=.1pt,draw=gray!20},}

\usepackage[normalem]{ulem}

\usepackage[acronym]{glossaries}
\newacronym{3d}{3D}{three dimensional}
\newacronym{aoa}{AoA}{angle of arrival}
\newacronym{aod}{AoD}{angle of departure}
\newacronym{ap}{AP}{access point}
\newacronym{awgn}{AWGN}{additive white gaussian noise}
\newacronym{b5g}{B5G}{Beyond-5G}
\newacronym{BSW}{BSW}{beam sweeping}
\newacronym{bsw}{CB-BSW}{codebook-based beam sweeping}
\newacronym[plural=BSs, firstplural=base stations (BSs)]{bs}{BS}{base station}
\newacronym{ctrl}{ctrl}{control}
\newacronym{cc}{CC}{control channel}
\newacronym{cdf}{CDF}{cumulative density function}
\newacronym{ce}{CE}{channel estimation}
\newacronym{csi}{CSI}{channel state information}
\newacronym{dc}{DC}{data channel}
\newacronym{dl}{DL}{downlink}
\newacronym{dft}{DFT}{discrete Fourier transform}
\newacronym{doa}{DoA}{direction-of-arrival}
\newacronym{emf}{EMF}{electromagnetic field}
\newacronym{em}{EM}{electromagnetic}
\newacronym{fp}{FP}{fractional program}
\newacronym{harq}{HARQ}{hybrid automatic repeat request}
\newacronym[plural=HRISs, firstplural=Hybrid Reconfigurable Intelligent Surfaces (HRISs)]{hris}{HRIS}{hybrid reconfigurable intelligent surface}
\newacronym{ibcc}{IB-C}{in-band control}
\newacronym{ibno}{IB-no}{IB-no}
\newacronym{ibwf}{IB-wf}{IB-wf}
\newacronym{ios}{IoS}{Internet-of-Surfaces}
\newacronym{iot}{IoT}{Internet-of-Things}
\newacronym{iid}{i.i.d.}{independently identically distributed}
\newacronym[plural=KPIs, firstplural=key performance indicators (KPIs)]{kpi}{KPI}{key performance indicator}
\newacronym{lf}{LF}{low frequency}
\newacronym{los}{LoS}{line-of-sight}
\newacronym{mac}{MAC}{medium access control}
\newacronym{mcs}{MCS}{modulation and coding scheme}
\newacronym{mimo}{MIMO}{multiple-input multiple-output}
\newacronym{miso}{MISO}{multiple-input single-output}
\newacronym{ml}{ML}{machine learning}
\newacronym{mmse}{MMSE}{minimum mean squared error}
\newacronym{mrt}{MRT}{maximum-ratio transmission}
\newacronym{mse}{MSE}{mean squared error}
\newacronym{nlos}{NLoS}{non-line-of-sight}
\newacronym{nr}{NR}{new radio}
\newacronym{obcc}{OB-C}{out-of-band control} 
\newacronym{oce}{OPT-CE}{optimization based on channel estimation}
\newacronym{ofdm}{OFDM}{orthogonal frequency-division multiplexing}
\newacronym{pdf}{pdf}{probability distribution function}
\newacronym{pla}{PLA}{planar linear array}
\newacronym{pap}{P\&P}{plug-and-play}
\newacronym{phy}{PHY}{physical}
\newacronym{pdcch}{PDCCH}{physical downlink control channel}
\newacronym{pucch}{PUCCH}{physical uplink control channel}
\newacronym{ppp}{PPP}{Poisson point process}
\newacronym{prb}{PRB}{physical resource block}
\newacronym{qos}{QoS}{Quality-of-Service}
\newacronym{rrc}{RRC}{radio resource control}
\newacronym[plural=RISs, firstplural=Reconfigurable intelligent surfaces (RISs)]{ris}{RIS}{reconfigurable intelligent surface}
\newacronym{risc}{RISC}{RIS controller}
\newacronym[plural=RFs]{rf}{RF}{radio frequency}
\newacronym{rmse}{RMSE}{root-mean-square error}
\newacronym{rss}{RSS}{received signal strength}
\newacronym{se}{SE}{spectral efficiency}
\newacronym{sdp}{SDP}{semidefinite programming}
\newacronym{sdr}{SDR}{semidefinite relaxation}
\newacronym{sinr}{SINR}{signal-to-interference-plus-noise ratio}
\newacronym{smse}{SMSE}{sum mean squared error}
\newacronym{snr}{SNR}{signal-to-noise ratio}
\newacronym{soa}{SoA}{state-of-the-art}
\newacronym{sre}{SRE}{smart radio environment}
\newacronym{ss}{SS}{synchronization signal}
\newacronym{toa}{ToA}{time-of-arrival}

\newacronym{tti}{TTI}{transmission time interval}
\newacronym[plural=UEs, firstplural=users equipment (UEs)]{ue}{UE}{user equipment}
\newacronym{ul}{UL}{uplink}
\newacronym{ula}{ULA}{uniform linear array}
\newacronym{urllc}{URLLC}{ultra-reliable low-latency communications}

\newacronym{fdd}{FDD}{frequency division multiplexing}
\newacronym{tdd}{TDD}{time division multiplexing}

\definecolor{amaranth}{rgb}{0.9, 0.17, 0.31}
\definecolor{cadmiumgreen}{rgb}{0.0, 0.42, 0.24}

\title{Control Plane for \\ Reconfigurable Intelligent Surfaces}

\author{
    Fabio Saggese,~\IEEEmembership{Member,~IEEE}, Victor Croisfelt,~\IEEEmembership{Student~Member,~IEEE}, \\
    Kyriakos Stylianopoulos,~\IEEEmembership{Student~Member,~IEEE}, George C. Alexandropoulos,~\IEEEmembership{Senior~Member,~IEEE},\\ and Petar Popovski,~\IEEEmembership{Fellow,~IEEE}\vspace{-1cm}
    \thanks{This work has been supported by the Villum Investigator grant ``WATER'' from the Villum Foundation, Denmark, and the SNS JU TERRAMETA under the EU's Horizon Europe research and innovation program under Grant Agreement No 101097101, including top-up funding by UKRI under the UK government's Horizon Europe funding guarantee.} 
    \thanks{F. Saggese, V. Croisfelt, and P. Popovski are with the Department of Electronic Systems, Aalborg University, Aalborg, Denmark (e-mails: \{fasa, vcr, petarp\}@es.aau.dk).} 
    \thanks{K. Stylianopoulos and G. C. Alexandropoulos are with the Department of Informatics and Telecommunications, National and Kapodistrian University of Athens, Panepistimiopolis Ilissia, 15784 Athens, Greece (e-mails: \{kstylianop, alexandg\}@di.uoa.gr).}
}

\begin{document}

\maketitle

\begin{abstract}
    Research on reconfigurable intelligent surfaces (RISs) has predominantly focused on purely physical (PHY)-layer aspects, particularly, on how signals are dynamically shaped by a controllable wireless propagation environment. However, integrating RISs as system-level network elements requires the development of an RIS-compatible control plane.
    In this article, we explore design options for such a control plane across two key dimensions: \emph{i}) the allocation of spectral resources for the control plane (in- or out-of-band), and \emph{ii}) the rate selection for the data plane (multiplexing or diversity). 
    While our analysis is necessarily simplified, it reveals the fundamental trade-offs inherent in these design choices, which are crucial for integrating RIS technology into future networks.
\end{abstract}

\section{Introduction}\label{sec:intro}
\Glspl{ris} are considered among the key components of the upcoming sixth generation (6G) of wireless networks due to their ability to dynamically shape the propagation environment~\cite{RISE6G_COMMAG_all,Wu2021:tutorial}. The majority of research efforts in \glspl{ris} focuses solely on \gls{phy}-layer aspects, such as: the development of realistic channel models accounting for the electromagnetic interactions among \gls{ris} elements, source, and destination~\cite{GR_003}; specific \gls{ce} methods for end-to-end or individual channels~\cite{Jian2022:hw-ce-tutorial} tailored to specific \gls{ris} hardware~\cite{RISsurvey2023}; and, enhancements to \gls{phy}-layer communication metrics, e.g., spectral and energy efficiency, by optimizing the \gls{ris} alone or jointly with the operations of the \gls{bs}~\cite{bjornson2022reconfigurable_all}. However, a significant challenge remains: how to effectively integrate \glspl{ris} into existing networks. Addressing this question requires careful consideration of the \gls{ris} deployment impact on both the control and data planes. Our study in this article concentrates on designing an \gls{ris}-compatible control plane and exploring the trade-offs associated with different design choices.

\subsection{Data and Control Planes}
The main goal of an \gls{ris} is to favorably tune the wireless propagation environment so that user data is transmitted reliably. We refer to the suite of operations dedicated to data transmission as the \emph{data plane}. However, to achieve the promised potential over the data plane, it is necessary to properly design the control procedures, protocol aspects, and signaling operations, altogether referred to as the \emph{control plane}. 
Despite its importance, there has been considerably less focus in the existing body of research concerning integrating \glspl{ris} at the system level compared to purely \gls{phy}-layer topics~\cite{ABoI_EURASIP}. Notably, there is a significant gap in the literature concerning the systematic examination of communication paradigms in the presence \glspl{ris} and their associated control plane aspects for integrating the technology into future networks.

\subsection{Communication Paradigms}
We can traditionally distinguish between two different communication paradigms: \emph{multiplexing-} and \emph{diversity-oriented}. The former is based on adapting a \gls{kpi}, such as data rate, to meet a \gls{qos} to the conditions of the propagation environment. On the other hand, a diversity-oriented setup defines a \gls{kpi} \emph{a priori}, and the devices expect the propagation environment to meet the \gls{qos} specifications; if those are not met, the system experiences a failure. Not surprisingly, the literature reveals that developed communication schemes for \gls{ris}-empowered networks commonly fall in one of the two paradigms. Many \gls{ris} optimization techniques are based on the knowledge of instantaneous \gls{csi}~\cite{Wu2021:tutorial}. To this end, \gls{csi} is acquired and fed as input to an optimization procedure -- e.g., to configure the tunable \gls{ris} elements response -- that, along with data transmission, needs to be finalized within one or a few multiples of the channel coherence time. We refer to these approaches as \gls{oce} schemes, which fall into the multiplexing-oriented paradigm. However, estimating the instantaneous \gls{csi} in the presence of \glspl{ris} is generally complex and generates overhead due to the large number of their passive elements~\cite{Jian2022:hw-ce-tutorial}. To avoid \gls{csi} acquisition and a subsequent complex optimization procedure, an alternative approach is the \gls{bsw}: the \gls{ris} cycles through different \emph{configurations} from an available list, each dictating the reflecting state of the \gls{ris} elements, letting the system find a setup suitable to satisfy a pre-determined \gls{qos}. This scheme follows the diversity-oriented paradigm~\cite{alexandropoulos2022hierarchical,An2022}.

\subsection{The RIS Control Challenge}
Crucially, some of the existing control signaling schemes between a \gls{bs} and \glspl{ue} employed in 5G \gls{nr} cannot be directly used for \gls{ris}-aided networking without considering their influence on the data plane~\cite{Lin2019:5gnr}. In this matter, addressing the availability of physical time-frequency resources utilized for control operations is imperative. This involves contemplating the coexistence of the control and data planes.
The main consideration is that, if the control is carried over a wireless channel, the influence of the \gls{ris} on the environment might, paradoxically, negatively impact control signaling, giving rise to a ``chicken-egg dilemma,'' where the control of the \gls{ris} is affected by the current state of the \gls{ris} itself. This is particularly important when control and data planes share the same spectral resources, constraining them from co-existing simultaneously to have reliable control, while possibly inducing unwanted communication overhead. Thus, we have to differentiate between two options: \textit{i}) \emph{\gls{ibcc}}, which is affected by the \gls{ris} configuration and/or shares the resources with the data plane; and \textit{ii}) \emph{\gls{obcc}}, which is not. Even more so, the response of the \gls{ris} to control is not immediate, which introduces further overhead that can be detrimental to performance.

In light of the afore-described challenges, understanding the possible options for the control plane is crucial to seamlessly integrate \glspl{ris} as novel network components into the current wireless infrastructure. To this end, in this article, we discuss potential control options for \gls{ris}-aided wireless networks, describing how to evaluate the RIS impact on communication performance and study the performance obtained by the \gls{oce} and \gls{bsw} schemes, used as extreme examples.

\section{RIS-Compatible Control Taxonomy}~\label{sec:taxonomy}
The radio interface of \gls{nr} comprises the \gls{phy} layer and higher layers, including \gls{mac} and \gls{rrc}~\cite{Lin2019:5gnr}. The \gls{nr} control plane encompasses all cross-layer aspects of the radio interface pertinent to control, being engineered to facilitate communications between network devices. For instance, physical channels like the \gls{pdcch} have the role of conveying control information from the \gls{bs} to the \gls{ue}, such as resource allocation, scheduling, and \gls{harq} feedback. However, when contemplating the creation of an \gls{ris}-compatible control plane, merely treating the \gls{ris} as a novel kind of \gls{ue} may prove inadequate depending on the context. This is because the \gls{ris} does not operate in isolation, but interacts with and influences the operation of all other wireless network devices sharing the same propagation environment. Hence, our discussion will address the control plane on two distinct fronts: for the \gls{ris} and the \glspl{ue}, expanding upon the established standard~\cite{Lin2019:5gnr}. 

On the other hand, it is important to note that the data plane may undergo smaller adjustments that are less critical compared to those in the control plane. This is because the goal of the data plane remains to be the communications between \gls{bs} and \glspl{ue}. However, the physical aspects of the data plane are subject to changes due to \gls{ris} control. Hence, there are new consequences of how the \gls{ris} control plane can interact with both the control plane of the \gls{ue} and the data plane. 

\begin{table*}[tbh]
    \centering
    \begin{tabular}{p{0.5\columnwidth}|p{0.55\columnwidth}p{0.55\columnwidth}}
        \toprule
        \centering \textbf{Implicit control} & \multicolumn{2}{c}{\textbf{Explicit control}} \\        
        \midrule
        \multirow{3}{0.5\columnwidth}{\justifying \emph{NO dedicated instructions} are sent toward the device. All decisions concerning device operations must be made locally by the device itself and can be based on other received and interpreted signals (e.g., pilot symbols, scheduling information, \glspl{ss}, etc.), which implicitly control the device's behavior.} & \multicolumn{2}{p{1.1\columnwidth}}{\emph{Explicit} instructions are sent to the device from a third-party decision-maker.} \\
        & \begin{center} \emph{Out-of-band Control} (OB-C) \end{center} & \begin{center} \emph{In-band Control} (IB-C) \end{center}\\
        & Control employs resources orthogonal to the \gls{ris} operational spectrum resources; control {DOES NOT influence the data plane} (simpler control design, lower overall spectral efficiency, and higher cost). & 
        Control employs resources overlapping the \gls{ris} operational spectrum resources; control DOES influence the data plane (more complex control design, higher spectral efficiency, and reduced cost). \\
        \bottomrule
    \end{tabular}
    \caption{Taxonomy of the \gls{ris}-compatible control plane.}
    \label{tab:kind-of-control}
\end{table*}

\subsection{Control for \glspl{ris}}
To remain open to various hardware implementations of the \gls{ris}~\cite{RISsurvey2023}, we abstract the \gls{ris} device as comprising two distinct modules pertaining to data and control planes, respectively: the \gls{ris} panel and the \gls{risc}. The former encompasses all hardware and software components related to the electronically controlled elements and their tuning based on inputs from the \gls{risc}; the latter comprises components capable of receiving and processing control signals while directing the behavior of the \gls{ris} panel accordingly. The design of the \gls{ris} control plane, responsible for managing control signaling between the \gls{ris} and the network, influences the processing capacity of the \gls{risc} and, in certain instances, the hardware capabilities of the overall \gls{ris} device. Additionally, the presence of the \gls{ris} may impact the design of the \gls{ue} control plane. In light of these considerations, a potential taxonomy of control planes for \glspl{ris} is outlined in Table~\ref{tab:kind-of-control}.

The control of the \gls{ris} can be \emph{implicit} or \emph{explicit}. In the presence of \emph{implicit control}, the control plane of the \gls{ris} does not have dedicated resources and specific instructions designed for the \gls{risc}, but rather exchange information sharing the control plane of the \glspl{ue}, e.g., through the \gls{pdcch} and \gls{pucch} in \gls{nr}. This detaches the control of the \gls{ris} from a possible third-party decision-maker, letting the \gls{risc} decide autonomously on the behavior of the \gls{ris} panel. Removing the burden of dedicated control operations comes at the cost of increased complexity regarding the \gls{risc}, which needs increased processing capabilities to infer the optimal control of the \gls{ris} panel. Moreover, the \gls{risc} generally needs to acquire local \gls{csi}~\cite{Jian2022:hw-ce-tutorial} or other channel statistics~\cite{HRIS_DoA} to perform its decision; therefore, many solutions in the literature require that the \gls{ris} panel is equipped with sensing capabilities~\cite{Alexandropoulos2024:hris-magazine}, i.e., at least a receiving \gls{rf} chain working on the same resources used by the data plane, increasing the manufacturing cost and creating the need of specific orchestration methods in \gls{phy} and \gls{mac} layers to avoid that the influence of the \gls{ris} decreases the communication performance instead of improving it; see, orchestration examples in~\cite{croisfelt2023orchestration_all}. 

In an \emph{explicit control} scenario, the \gls{risc} receives dedicated control information to command the behavior of the \gls{ris} panel. This implies that a third-party device, e.g., the \gls{bs}, acts as the network's decision-maker communicating with the \gls{risc}. Consequently, explicit control necessitates that the \gls{risc} is equipped with a dedicated transceiver for control signaling, which may be conveyed over \glspl{rf} or wired connections, depending on the specific environmental and infrastructural needs.
While this approach reduces the processing burden of the \gls{risc}, it also increases the complexity of control operations and their impact on communication performance. 
We can further distinguish two different kinds of explicit control depending on the available \gls{phy} resources: 
\begin{itemize}
    \item \emph{\gls{obcc}}, in which the physical resources used by the control plane are orthogonal to those of the data plane. It can represent either the case in which a portion of the spectrum is reserved for control signaling if the \gls{risc} communicate with the other device wirelessly or the case of a dedicated cable connection between the decision-maker and the \gls{risc}.
    \item \emph{\gls{ibcc}}, in which the physical resources are shared between control and data planes, implicitly assuming a wireless communication is needed between the \gls{risc} and the decision-maker.
\end{itemize}

Finally, we remark that \emph{hybrid modes} for \gls{ris} control are possible. For example, an \gls{ris} can act in an autonomous way (implicit control) until it receives instructions to load a specific configuration from the \gls{bs} (explicit control) because of the particular needs of the system in that instant. In this case, the technical requirements of implicit and explicit control must be satisfied simultaneously by the \gls{ris} hardware.

Henceforth, we will analyze the two extremes of \gls{obcc} and \gls{ibcc} to better understand their trade-offs. Future research should focus on enabling hybrid control of \glspl{ris}.

\begin{figure}
    \centering
    \begin{subfigure}{\columnwidth}
        \centering
        \includegraphics[width=.9\textwidth]{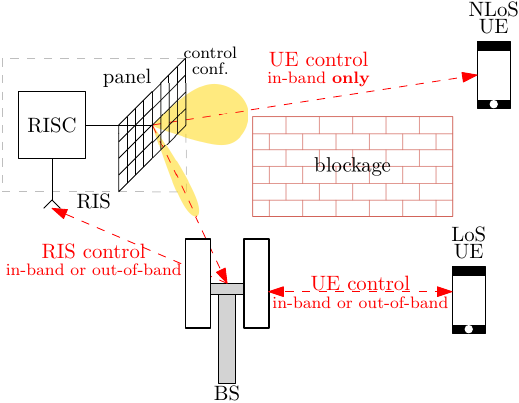}
        \caption{Control plane.}
    \end{subfigure}  
    \begin{subfigure}{\columnwidth}
        \centering
        \includegraphics[width=.9\textwidth]{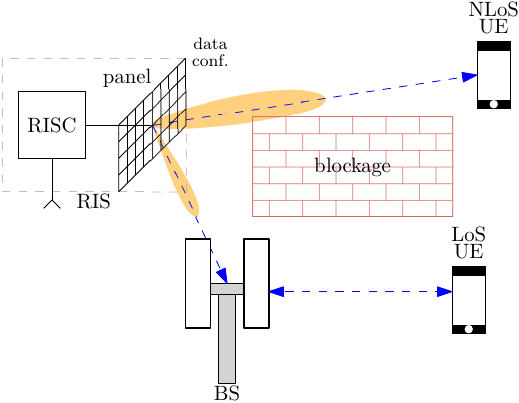}
        \caption{Data plane.}
    \end{subfigure}
    \caption{
        Toy example of data and control planes for explicit control, featuring one \gls{ue} in a \gls{los} condition and another one in a \gls{nlos} condition relative to the \gls{bs}. ``Conf.'' abbreviates the configuration of the \gls{ris} response/reflection.
        }
    \label{fig:control-data-plane}
\end{figure}

\subsection{Control for \glspl{ue}}
In the absence of an \gls{ris}, the control signaling to and from \glspl{ue} may also occur through either \gls{ibcc} or \gls{obcc}. On the other hand, the deployment of an \gls{ris} may constrain the kind of \gls{ue} control employed. As depicted in Fig.~\ref{fig:control-data-plane}, if the \gls{ue} has a direct link with the \gls{bs}, the kind of control is not constrained in any way since the \gls{bs}-\gls{ue} control signaling can occur as usual. However, when a \gls{ue} has a direct link with the \gls{ris} only, the \gls{bs}-\gls{ue} control signals need to impinge on the \gls{ris} panel to reach the destination. Considering that the \gls{ris} panel is designed to work on the data plane resources, the control is constrained to be an \gls{ibcc}. In turn, this generates the need for a specific configuration loaded by the \gls{ris} to convey control information to the area not covered by the \gls{bs}. This \emph{\gls{ctrl} configuration} must be designed to provide reliable communication performance for a wide area. Potential candidates for \gls{ctrl} configurations are wide-width beams, as in~\cite{Ramezani2023:ris-broad}, and hierarchical beam codebooks, offering lower latency beam tracking as well as adjustable reliability and coverage levels~\cite{alexandropoulos2022hierarchical}.

\section{Evaluating the Impact of RIS Control}\label{sec:signaling}
To explore the implications of an \gls{ris}-compatible control plane using the introduced taxonomy, we begin by abstracting the concept of a communication protocol into its fundamental phases, or ``atoms.'' This abstraction enables us to analyze how the control plane influences specific protocols varying the number and arrangement of these ``atoms.'' We then utilize this framework to examine how an \gls{ris}-compatible control plane affects two common instances of communication paradigms for \gls{ris}-aided systems: \gls{oce} and \gls{bsw}.

\subsection{Abstract View of a Communication Protocol}\label{subsec:abstract}
In a generic wireless network, communication happens according to a frame-based protocol, whose boundaries are imposed by the control information dictating the frame length and structure. The typical way communication happens can be broken down into three main phases, possibly occurring more than once within each frame: ``\emph{Payload},'' ``\emph{Algorithmic},'' and ``\emph{Signaling}.'' A certain amount of time, spectral, and computational resources are reserved for each phase, based on the system requirements.

\paragraph*{Payload} Payload phases are tied with the data plane, representing the portion of the frame used for the actual data transmission. At the \gls{phy} layer, the amount of time and spectral resources employed by this phase determine the achievable communication performance in terms of data rate and reliability.

\paragraph*{Algorithmic} Algorithmic phases involve the processing operations aimed to infer, optimize, and set the transmission parameters, comprehending operations as \gls{csi} acquisition, \gls{bs} and/or \gls{ue} beamforming, \gls{ue} spectral/time scheduling, selection of \gls{mcs}, etc. The specifics of this phase depend on the chosen communication paradigm and the computational complexity of the solution employed. The time resources reserved for the Algorithmic phases contribute to the overhead time; moreover, eventual errors in any algorithmic task at hand may result in an unexpected decline in communication performance. Given the total available resources and the computational capability of the devices involved, we can generally state that the larger the number of resources reserved for an Algorithmic phase, the fewer errors in the optimization process, e.g., transmitting a higher number of pilot sequences improves \gls{csi} acquisition, the number of gradient-descent iterations contributes to the optimality of the solution, etc. At the same time, fewer resources are left available for a Payload phase, resulting in lower communication performance. Hence, there exists a trade-off between the resources given for Algorithmic phases and the communication performance.

\paragraph*{Signaling} Within the Signaling phases, control messages are exchanged with the network nodes to inform them of the frame structure, to command the algorithmic operations, or to instruct the setting of the communication parameters in preparation for a Payload phase. Resources reserved for Signaling phases may introduce overhead when exerting \gls{ibcc}, thus decreasing the overall communication performance. On the other hand, given the information content of the signaling, the amount of resources allocated for the transmission dictates the reliability of the control messages. Again, there exists a trade-off between the resources reserved for the Signaling and the Payload phases.

\subsection{The Causality Problem}
An important consideration regards the \emph{causality} of the Algorithmic and Signaling phases. The information regarding the communication parameters conveyed in the control messages needs to be decoded before the data is. This generates limited issues when considering fully digital hardware, which can decide the order of information decoding after their reception. Conversely, operation order is of fundamental importance when analog hardware is considered. Let us imagine a device equipped with a hybrid digital and analog beamformer receiving data from the \gls{bs}, the latter being the decision-maker of the network. While the digital processing of the received data can be performed after its reception, impacting the end-to-end latency, the analog part of the device beamformer must be loaded before the transmission takes place, otherwise, the data might be impossible to decode. Therefore, a control message informing the device of the beamformer to load needs to be sent \emph{prior to} the data transmission, which, in turn, constrains the Algorithmic phase to output this parameter before the relevant Signaling phase. Remarking that the \gls{ris} panel is fundamentally analog hardware, this poses a fundamental issue to consider when designing the control plane for \gls{ris}-aided wireless networks.

\subsection{Control Procedures for RIS-Empowered Systems}
\begin{figure}[!t]
    \centering
    \begin{subfigure}[t]{0.99\columnwidth}
        \centering
        \includegraphics[width=0.99\textwidth]{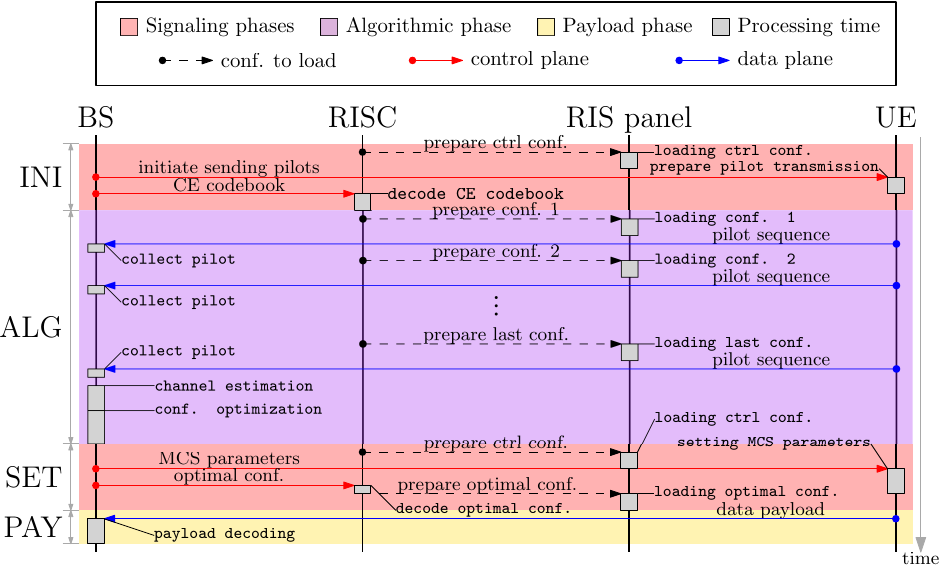}
        \caption{\gls{oce} operations.}
        \label{fig:RIS-oce}
    \end{subfigure}\vspace{0.3cm}
    \begin{subfigure}[t]{0.99\columnwidth}
        \centering
        \includegraphics[width=0.99\textwidth]{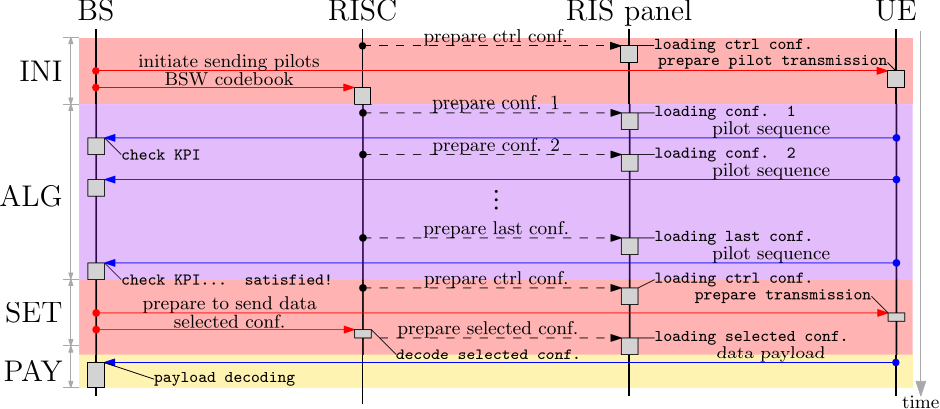}
        \caption{\gls{bsw} operations.}
        \label{fig:RIS-bsw}
    \end{subfigure} 
    \caption{Control information exchange diagram of the \gls{oce} and \gls{bsw} schemes. \texttt{Monospace font} is used to indicate processing operations. ``Conf.'' abbreviates the configuration of the RIS response.}
    \label{fig:controlled-frames}
\end{figure}

We will now focus on the description of the minimum control procedures needed for an \gls{ul} communication within an \gls{ris}-empowered system employing \gls{tdd} and explicit control, showing the difference between the \gls{oce} and \gls{bsw} schemes. The scenario at hand focuses on a single \gls{ue} not having a direct link with the \gls{bs}, necessitating an \gls{ris} to extend the coverage of the network to highlight the aforementioned special requirements of the control plane -- see the \gls{nlos} \gls{ue} in Fig.~\ref{fig:control-data-plane}. 
According to the consideration on \gls{ue} control, it is assumed that the \gls{ris} loads the \gls{ctrl} configuration at any time it is in an idle state, i.e., anytime it does not receive a command to load a specific configuration. In this case, we observe that the typical structure includes a single Payload phase (PAY), a single Algorithmic (ALG) phase, and two Signaling phases: one placed at the beginning of the frame to initialize the algorithmic operation, called ``Initialization (INI),'' and the other following the Algorithm phase to set the communication parameter for the transmission, called ``Setup (SET).'' An illustrative example of the diagram of the control and data plane operations and their order is depicted in Fig.~\ref{fig:controlled-frames}. We remark that an optimization of the timing of the operations can be done through pipelining of various processing times.
Finally, we highlight that our approach can be easily adapted to describe other protocols, for example, incorporating additional signaling phases to accommodate feedback in \gls{fdd} systems.

\subsubsection{Control with Channel Estimation}
The main idea of the \gls{oce} approach is to adapt the transmission data rate according to the propagation conditions. Therefore, the \gls{bs} aims to acquire the instantaneous \gls{csi} of the \gls{ue}-\gls{ris}-\gls{bs} end-to-end channel and use it to optimize the \gls{ris} configuration and the \gls{mcs} according to the \gls{ue} target \gls{qos}. Generally speaking, \gls{csi} acquisition relies on transmission of replicas of known pilot sequences, while the \gls{ris} sweeps through a \emph{\gls{ce} codebook of configurations}, one per replica, to solve the indeterminacy of the \gls{ue}-\gls{ris} and \gls{ris}-\gls{bs} paths~\cite{Jian2022:hw-ce-tutorial}. On the other hand, the optimization of network parameters depends on the goal of the system; for example, to maximize the \gls{ue} data rate, the \gls{ris} configuration-induced phase shifts should compensate the phase of the \gls{ue}-\gls{ris}-\gls{bs} end-to-end channel, and the \gls{mcs} should be set to the maximum achievable under the optimal configuration~\cite{bjornson2022reconfigurable_all}.

Following the diagram of Fig.~\ref{fig:RIS-oce}, we can detail the control operations needed. During the INI phase, while the \gls{ris} has loaded the \gls{ctrl} configuration, the \gls{bs} informs the \gls{ue} and the \gls{risc} that the \gls{ce} procedure is starting, hence informing the latter about the employed \gls{ce} codebook, and the former about the pilot sequences to use, their number of replicas, and the periodicity of pilot transmission, taking into account the time the \gls{risc} needs to load each configuration from the codebook. Then, the ALG phase comprises the reception of the pilot sequences, the acquisition of the \gls{csi} from the received data, and the optimization of the \gls{ris} configuration and \gls{mcs}~\cite{bjornson2022reconfigurable_all}. After that, the SET phase starts: the \gls{risc} loads the \gls{ctrl} configuration and the \gls{bs} instructs the \gls{ue} about the \gls{mcs} to use for data transmission while informing the \gls{ris} about the optimal configuration to load. Finally, the PAY phase occurs.

We remark that the explicit signaling of a configuration toward the \gls{risc} might require the transmission of a relatively large amount of control data, because of the need to transmit the phase shift value for each element of the \gls{ris} panel. While a single tunable element is generally controlled by $1$ or $2$ bits only, the number of elements tends to be very high to provide substantial gain to the reflected signal~\cite{Wu2021:tutorial}. Moreover, the number of configurations in the \gls{ce} codebook is on the order of the number of \gls{ris} panel's elements to solve the indeterminacy of \gls{csi} acquisition~\cite{Jian2022:hw-ce-tutorial}. Accordingly, the control plane is easily susceptible to \emph{control errors}, i.e., errors in the data plane -- e.g., communication outages~\cite{saggese2023impact} -- generated by the incorrect decoding of control information, if the resource orchestration is not specifically designed to provide reliable transmission of a large amount of control data.

\subsubsection{Control with Beam Sweeping}
In \gls{bsw}, the \gls{bs} chooses a communication \gls{kpi}, e.g., the communication \gls{snr}, prior to the transmission and seeks for a suitable \gls{ris} configuration among a pre-determinate set of configurations, namely the \emph{BSW codebook}, achieving a target \gls{qos}. Like the previous communication approach, the \gls{ris} switches through the codebook while the \gls{ue} transmits pilot sequences. Then, the \gls{bs} evaluates the \gls{kpi} for each pilot sequence and selects the most suitable configuration among the ones satisfying the aimed \gls{qos}. The communication fails if no configuration satisfies it and an outage occurs.

Following the diagram of Fig.~\ref{fig:RIS-bsw}, we can detail the control operations needed for \gls{bsw}. During the INI phase, while the \gls{ris} has the \gls{ctrl} configuration loaded, the \gls{bs} informs the \gls{risc} about the BSW codebook and the \gls{ue} about the pilot signaling; this needs similar control information to \gls{oce}. Then, the ALG phase comprises the reception of the pilot sequences and the evaluation of the \gls{kpi}~\cite{An2022}. During the SET phase, with the \gls{risc} loading the \gls{ctrl} configuration, the \gls{bs} informs the \gls{ue} to prepare data transmission. Then, the \gls{bs} commands the \gls{risc} to load the selected configuration. Finally, the PAY phase takes place. 

We note that the beam sweeping process during the ALG phase may allow stopping the procedure as soon as a \gls{kpi} value measured is above the target \gls{qos}, instead of waiting to check the whole BSW codebook; this reduces the overhead time. We refer to this as \emph{early stopping}. However, this constrains the control plane design to be able to promptly inform the \gls{ue} about the sudden end of the ALG phase, thus reserving resources for eventual SET control message transmission after \emph{each} \gls{kpi} evaluation.

Finally, it is worth mentioning that signaling to the \gls{risc} to load a configuration from a codebook requires fewer control data than the one needed when specifying the phase shift of each \gls{ris} element. Indeed, transmitting the integer index of the selected configuration in the BSW codebook is enough to instruct the \gls{risc} unambiguously.

\section{Performance Evaluation and Comparison}\label{sec:comparison}
In this section, the performance evaluation of the system under consideration is presented\footnote{The simulation code and detailed parameters can be found at \url{https://github.com/lostinafro/ris-control}.}. Results are presented through numerical simulations of an \gls{ris}-empowered system employing narrowband data transmission, in which a single \gls{nlos} \gls{ue} is served by the \gls{bs} through the \gls{ris}, resulting in an \gls{ibcc} for the \gls{ue} (see Fig.~\ref{fig:control-data-plane}). Regarding \gls{ris} control, the performance is evaluated for both \gls{obcc} and \gls{ibcc} schemes. Moreover, the frame structure has been organized to work with quantities resembling 5G \gls{nr} standard, i.e., single \gls{prb} \gls{ofdm} transmission with numerology $0$ in which the basic \gls{tti} for control and data operations is set as the half of a subframe, i.e., $0.5$ ms~\cite{saggese2023impact}.

\begin{figure}[!t]
    \centering
\begin{tikzpicture}

\def\hsep{2.5cm}
\def\vsep{0.8cm}
\def\vside{3.5cm}

\begin{groupplot}[
group style={group name=tau, group size=1 by 2,  horizontal sep=\hsep, vertical sep=\vsep, xlabels at=edge bottom}, 
title style={at={(0.5,0.85)}},
anchor=south east,
height=\vside,
width=0.9\columnwidth,
xmajorgrids,
ymajorgrids,
legend cell align={left},
legend style={  
  at={(0.5, 1.1)}, 
  draw=none,
  fill opacity=0,
  anchor=south,  
},
legend columns=3,
xlabel={Frame length [ms]},
xmin=10, xmax=100,
ymin=-0.09, ymax=1.1,
ytick={-0.5,0,0.5,1,1.5,2},
yticklabels={
  \(\displaystyle {\ensuremath{-}0.5}\),
  \(\displaystyle {0.0}\),
  \(\displaystyle {0.5}\),
  \(\displaystyle {1.0}\),
  \(\displaystyle {1.5}\),
  \(\displaystyle {2.0}\)
},
ylabel={Goodput [Mbps]},
]
\nextgroupplot[title=\gls{obcc}]
\addlegendimage{semithick, black}
\addlegendentry{\gls{oce}~~}
\addlegendimage{semithick, blue, dashed, mark=*, mark options={fill=white, solid}}
\addlegendentry{\gls{bsw}~~}
\addlegendimage{semithick, red, dotted, mark=triangle*, mark options={fill=white, solid}}
\addlegendentry{\gls{bsw}: early stopping}

\node [anchor=south, font=\scriptsize, align=center] at (39, 0.65) {\gls{bsw} \\ dominant};
\node [name=trade, anchor=south, font=\scriptsize, align=center] at (59, 0.75) {trade-off};
\draw[gray, dotted, semithick] (59,-1)--(59, 2);
\draw[->] (trade.west) -- ([xshift=-2mm]trade.west);
\draw[->] (trade.east) -- ([xshift=2mm]trade.east);
\node [anchor=south, font=\scriptsize, align=center] at (79, 0.65) {\gls{oce} \\ dominant};
\pattern [pattern=north east lines] (0,-1) -- (15,-1) -- (15,2) -- (0,2);
\node[name=null, rotate=90, align=center, font=\scriptsize, anchor=center] at (22, 0.5) {null-rate};
\draw[->] (null.north) -- ([xshift=-2mm]null.north);

\addplot [semithick, black]
table {%
10 0
15 0
20 0
25 0
30 0
35 0
40 0
45 0
50 0
55 0.0631643479447905
60 0.250902826558474
65 0.409758462308514
70 0.545920435808548
75 0.663927479508577
80 0.767183642746103
85 0.858292022073332
90 0.939277248141979
95 1.01173771357182
100 1.07695213245868
105 1.1359556543087
110 1.18959521962689
115 1.23857047491742
120 1.28346445893373
125 1.32476692422874
130 1.36289227680875
135 1.39819352919765
140 1.43097326355877
145 1.46149232658464
150 1.48997678540878
155 1.51662353721202
160 1.54160486702755
165 1.56507217685426
170 1.58715905669116
175 1.60798382910881
180 1.62765166972548
185 1.64625638382234
190 1.66388190244041
195 1.68060354830883
200 1.69648911188383
};
\addplot [semithick, blue, dashed, mark=*, mark options={fill=white, solid}, mark repeat=5, mark phase=0]
table {%
10 0
15 0
20 0.00472185126780811
25 0.0132211835498627
30 0.120546645844065
35 0.146378069953508
40 0.16575163803559
45 0.180819968766098
50 0.192874633350504
55 0.202737540737746
60 0.210956630227114
65 0.217911244410426
70 0.223872342281836
75 0.229038627103724
80 0.233559126322876
85 0.237547802104482
90 0.24109329168813
95 0.244265571841922
100 0.247120623980334
105 0.249703766391278
110 0.252052077673955
115 0.254196187975529
120 0.256161622418639
125 0.2579698221063
130 0.259638929510294
135 0.261184399328808
140 0.262619478445999
145 0.263955586589591
150 0.265202620856944
155 0.266369201300596
160 0.26746287046652
165 0.268490256652691
170 0.269457208357322
175 0.270368905678832
180 0.271229953149147
185 0.272044457512958
190 0.272816093226042
195 0.273548157876917
200 0.274243619295248
};
\addplot [semithick, red, dotted, mark=triangle*, mark options={fill=white, solid}, mark repeat=5, mark phase=5]
table {%
10 0.00402440224596746
15 0.00899388174393452
20 0.0167730282431831
25 0.0326220919463067
30 0.0434469668598781
35 0.0602712354156992
40 0.0767701106310641
45 0.0965329736917895
50 0.115759378457731
55 0.137666460796888
60 0.158073540454514
65 0.180703568346782
70 0.201779768770494
75 0.224925582631613
80 0.246063338852169
85 0.266643561831439
90 0.285222827753042
95 0.302629493907175
100 0.318474224074071
105 0.332887500766535
110 0.345998706534284
115 0.357969807452664
120 0.368943316627845
125 0.379038945069012
130 0.388357986707012
135 0.39698672896442
140 0.404999132489155
145 0.412458956460461
150 0.419421458833679
155 0.4259347675054
160 0.432040994385138
165 0.437777146908528
170 0.443175878695248
175 0.448266111522727
180 0.453073553637569
185 0.457621134016473
190 0.461929368059645
195 0.46601666702368
200 0.469899601039514
};

\nextgroupplot[title=\gls{ibcc}]

\node [anchor=south, font=\scriptsize, align=center] at (41, 0.65) {\gls{bsw} \\ dominant};
\node [name=trade, anchor=south, font=\scriptsize, align=center] at (61, 0.75) {trade-off};
\draw[gray, dotted, semithick] (61,-1)--(61, 2);
\draw[->] (trade.west) -- ([xshift=-2mm]trade.west);
\draw[->] (trade.east) -- ([xshift=2mm]trade.east);
\node [anchor=south, font=\scriptsize, align=center] at (81, 0.65) {\gls{oce} \\ dominant};
\pattern [pattern=north east lines] (0,-1) -- (17,-1) -- (17,2) -- (0,2);
\node[name=null, rotate=90, align=center, font=\scriptsize, anchor=center] at (24, 0.5) {null-rate};
\draw[->] (null.north) -- ([xshift=-2mm]null.north);

\addplot [semithick, black]
table {%
10 0
15 0
20 0
25 0
30 0
35 0
40 0
45 0
50 0
55 0
60 0.173701956848174
65 0.338496121037468
70 0.479748261771148
75 0.602166783740337
80 0.709282990463378
85 0.80379729051312
90 0.887810001668446
95 0.962979269544264
100 1.0306316106325
105 1.0918408716171
110 1.14748565433036
115 1.19829176028596
120 1.24486402407858
125 1.2877105067678
130 1.32726110617323
135 1.36388203154863
140 1.39788717654007
145 1.42954713911831
150 1.45909643752466
155 1.48673932958222
160 1.51265454088618
165 1.53699913332324
170 1.55991169091106
175 1.58151495949385
180 1.60191804648872
185 1.62121826391629
190 1.63950268042663
195 1.65684943455182
200 1.67332885097075
};
\addplot [semithick, blue, dashed, mark=*, mark options={fill=white, solid}, mark repeat=5, mark phase=0]
table {%
10 0
15 0
20 0
25 0.00944370253561621
30 0.100455538203388
35 0.129157120547213
40 0.150683307305082
45 0.167425897005646
50 0.180819968766098
55 0.191778754751922
60 0.200911076406775
65 0.208638425499344
70 0.215261867578688
75 0.221002184047453
80 0.226024960957622
85 0.230456822937184
90 0.234396255807905
95 0.237921011534339
100 0.24109329168813
105 0.243963449922513
110 0.246572684681043
115 0.248955029460569
120 0.251138845508469
125 0.253147956272537
130 0.255002520054753
135 0.256719708741991
140 0.258314241094426
145 0.259798805698417
150 0.261184399328808
155 0.262480599821755
160 0.263695787783893
165 0.264837327990749
170 0.265911718773673
175 0.266924715797573
180 0.267881435209034
185 0.268786440057713
190 0.269643813072251
195 0.27045721823989
200 0.271229953149147
};
\addplot [semithick, red, dotted, mark=triangle*, mark options={fill=white, solid}, mark repeat=5, mark phase=5]
table {%
10 0
15 0.00644964362683847
20 0.0130974622677949
25 0.0257648668862417
30 0.0357580044167536
35 0.0505661696450303
40 0.0661625029849401
45 0.0845606191892698
50 0.103231643456592
55 0.124412834007819
60 0.144549338209927
65 0.166518818849299
70 0.18757073769743
75 0.210033877409134
80 0.231809194186248
85 0.252511909085859
90 0.271849872008153
95 0.289586881333593
100 0.306083742129169
105 0.321052566906402
110 0.334701724213248
115 0.34716399827602
120 0.358587749500228
125 0.3690976006265
130 0.378799001666135
135 0.387781780406538
140 0.396122932094055
145 0.403888831941054
150 0.411137005131586
155 0.417917554245309
160 0.424274319039425
165 0.43024582536117
170 0.435866066605166
175 0.441165151206647
180 0.446169842219158
185 0.450904009393154
190 0.455389009873782
195 0.459644010329762
200 0.463686260762943
};
\end{groupplot}

\end{tikzpicture}          
    \caption{Analysis of the goodput performance vs. the frame length.}
    \label{fig:goodput}  
\end{figure}
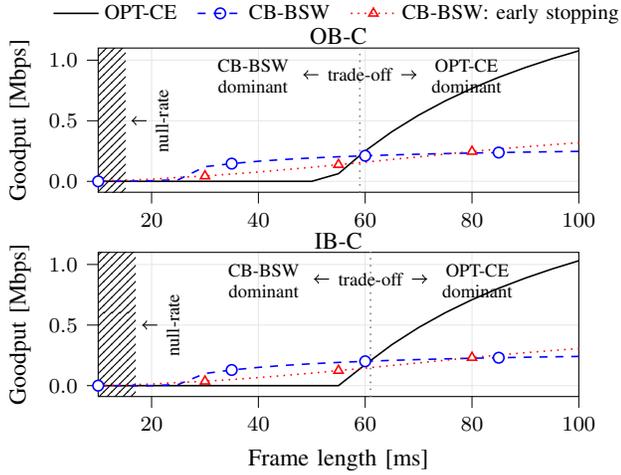

First, we show the results accounting for the overhead generated by the control plane, i.e., by the Signaling and Algorithmic phases, assuming perfect control reliability. To do so, we consider the goodput as the comparison metric, evaluating it for \gls{oce}, \gls{bsw}, and \gls{bsw} with early stopping. For \gls{bsw}, we consider the \gls{snr} as the \gls{kpi} under test, with target \gls{qos} of 10 dB. Fig.~\ref{fig:goodput} illustrates the average goodput as a function of the overall frame duration. It is worth noting that the frame duration is directly tied to the coherence time of the channel: the higher the coherence time, the higher the frame duration. The results show that the different controls impact the overhead negligibly ($\le 2$ ms) and, hence, the goodput performance. On the other hand, the choice of the communication approach is highly dependent on the frame time. When the frame duration is low, there is not enough time to perform the control operations, which results in a null rate. The main advantage of \gls{bsw} is the possibility of providing a non-null transmission rate even in the presence of a small coherence block. With \gls{obcc}, \gls{bsw} with early stopping is able to provide a non-null rate for frame duration $> 15$ ms, \gls{bsw} needs at least $25$ ms, while the \gls{oce} needs more time to obtain the \gls{csi} and perform the Payload phase ($\ge 50$ ms). Nevertheless, if the time horizon lasts for at least $59$ ms, the \gls{oce} outperforms the other two.

\begin{figure}[!t]
    \centering
    \begin{tikzpicture}

\def\width{.6\columnwidth}

\definecolor{redstart}{rgb}{0.945098039215686,0.411764705882353,0.0745098039215686}
\definecolor{obcolor}{rgb}{0.0, 0.72, 0.92}

\begin{groupplot}[group style={group name=obcc, group size=1 by 2, xlabels at=edge bottom, yticklabels at=edge left, vertical sep =1cm},
width=\width,
height=\width,scale only axis,
title style={at={(0.5,0.95)}},
tick align=outside,
xlabel={Avg. SNR \gls{ris}-\gls{cc} [dB]},
xmin=0, xmax=30,
xtick={0,10,20,30,40},
xticklabels={
  \(\displaystyle {0}\),
  \(\displaystyle {10}\),
  \(\displaystyle {20}\),
  \(\displaystyle {30}\),
  \(\displaystyle {40}\)
},
ylabel={Avg. SNR \gls{ue}-\gls{cc} [dB]},
ymin=0, ymax=30,
ytick={0,10,20,30,40},
yticklabels={
  \(\displaystyle {0}\),
  \(\displaystyle {10}\),
  \(\displaystyle {20}\),
  \(\displaystyle {30}\),
  \(\displaystyle {40}\)
},
]
\nextgroupplot[title={\gls{oce}}]
\tikzfading [name=sfuma, bottom color=transparent, top color=white, middle color=white]
\fill [obcolor, path fading=sfuma] (0,10.5) -- (30,10.5) -- (30,30) -- (0,30);
\node [align=center, anchor=center, font=\small, color=obcolor!60!black] at (3.5, 15) {\contour{white}{OB-C}};

\draw[<->, gray, semithick] (axis cs:2,0)--(axis cs:2,10.5);
\node [align=center, anchor=west, font=\scriptsize] at (axis cs:2,6) {$\approx 10.5$ dB};

\addplot graphics [includegraphics cmd=\pgfimage, xmin=0, xmax=30, ymin=0, ymax=30] {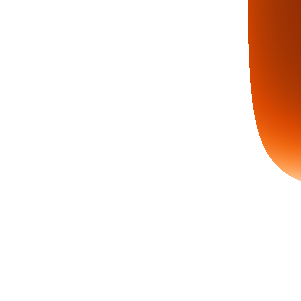};
\node [align=center, anchor=center, font=\small, color=redstart] at (27.5, 28) {\contour{white}{IB-C}};

\draw[<->, gray, semithick] (axis cs:0,26)--(axis cs:24.8,26);
\node [align=center, anchor=north, font=\scriptsize] at (axis cs:12.8,26.2) {$\approx 24.8$ dB};

\draw[<->, gray, semithick] (axis cs:29.5,0)--(axis cs:29.5,12);
\node [align=center, anchor=east, font=\scriptsize] at (axis cs:29.8,6) {$\approx 12$ dB};

\nextgroupplot[title={\gls{bsw}}]
\fill [obcolor, path fading=sfuma] (0,10.5) -- (30,10.5) -- (30,30) -- (0,30);
\node [align=center, anchor=center, font=\small, color=obcolor!60!black] at (3.5, 15) {\contour{white}{OB-C}};
\draw[<->, gray, semithick] (axis cs:2,0)--(axis cs:2,10.5);
\node [align=center, anchor=west, font=\scriptsize] at (axis cs:2,6) {$\approx 10.5$ dB};

\addplot graphics [includegraphics cmd=\pgfimage,xmin=0, xmax=30, ymin=0, ymax=30] {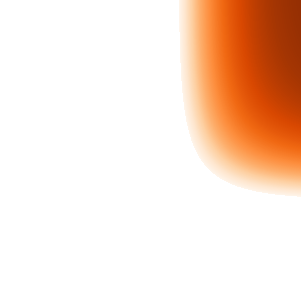};
\node [align=center, anchor=center, font=\small, color=redstart] at (27.5, 28) {\contour{white}{IB-C}};

\draw[<->, gray, semithick] (axis cs:0,26)--(axis cs:18,26);
\node [align=center, anchor=north, font=\scriptsize] at (axis cs:9,26.2) {$\approx 18$ dB};

\draw[<->, gray, semithick] (axis cs:29.5,0)--(axis cs:29.5,10.5);
\node [align=center, anchor=east, font=\scriptsize] at (axis cs:29.8,6) {$\approx 10.5$ dB};

\end{groupplot}

\begin{axis}[
        width=\width,
        height=\width,
        hide axis,
        point meta min=0.99, 
        point meta max=1,
        colorbar horizontal,
        colorbar style={
            at={(0,1.81)}, 
            anchor=south west,            
            tick pos=left,
            ticklabel pos=left,
            height=0.3cm,
            width=\width,
            xtick={0.99,0.992,0.994,0.996,0.998,1},
            xticklabels={
              \(\displaystyle {99.0}\),
              \(\displaystyle {99.2}\),
              \(\displaystyle {99.4}\),
              \(\displaystyle {99.6}\),
              \(\displaystyle {99.8}\),
              \(\displaystyle {100}\)
            },            
            title={{\footnotesize OB-C reliability \%}},
            title style={at={(0,0)}, anchor=east},
            },
        colormap={ibc}{[1pt]
          RGB(0pt)=(243, 252, 254);           
          rgb(8pt)=(0.0, 0.72, 0.92)
        },
    ]
    \end{axis}

    \begin{axis}[
        width=\width,
        height=\width,
        hide axis,
        point meta min=0.99, 
        point meta max=1,
        colorbar horizontal,
        colorbar style={
            at={(0,1.55)}, 
            anchor=south west,            
            tick pos=right,            
            height=0.3cm,
            width=\width, 
            xticklabels=\empty,
            title={{\footnotesize IB-C reliability \%}},
            title style={at={(0,0)}, anchor=east},
            },
            colormap={obc}{[1pt]
              rgb(0pt)=(1,0.96078431372549,0.92156862745098);
              rgb(1pt)=(0.996078431372549,0.901960784313726,0.807843137254902);
              rgb(2pt)=(0.992156862745098,0.815686274509804,0.635294117647059);
              rgb(3pt)=(0.992156862745098,0.682352941176471,0.419607843137255);
              rgb(4pt)=(0.992156862745098,0.552941176470588,0.235294117647059);
              rgb(5pt)=(0.945098039215686,0.411764705882353,0.0745098039215686);
              rgb(6pt)=(0.850980392156863,0.282352941176471,0.00392156862745098);
              rgb(7pt)=(0.650980392156863,0.211764705882353,0.0117647058823529);
              rgb(8pt)=(0.498039215686275,0.152941176470588,0.0156862745098039)
            },
    ]
    \end{axis}
\end{tikzpicture}     
    \caption{Evaluation of the reliability performance for \gls{obcc} and \gls{ibcc}. Reliability lower than 99\% is not colored.}
    \label{fig:reliability}    
\end{figure}
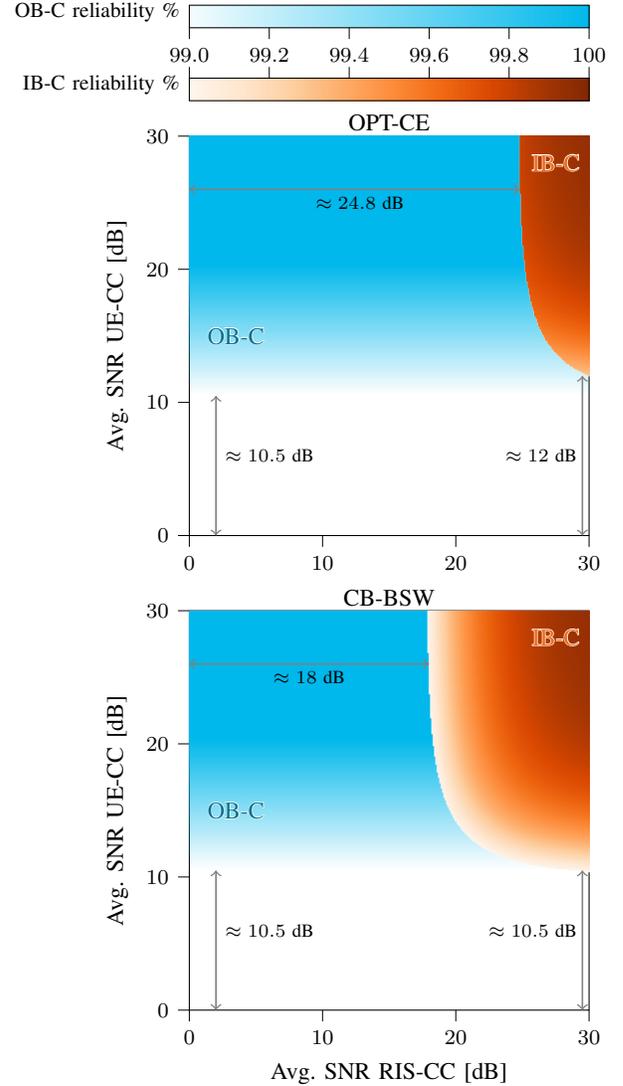

We now investigate the impact of the reliability of the control plane, comparing \gls{obcc} and \gls{ibcc} options. In the latter case, the \gls{phy} \gls{cc} connecting the \gls{bs} to the \gls{ue} (\gls{ue}-\gls{cc}) and to the \gls{risc} (\gls{ris}-\gls{cc}) are modeled as Rayleigh fading channels. When employing \gls{obcc}, the \gls{ue}-\gls{cc} is still modeled as a Rayleigh fading channel, while the \gls{ris}-\gls{cc} is an idealized error-free channel since the system designer can easily make it as reliable as possible.
To compare the two communication approaches, we evaluate the minimum average \gls{snr} providing at least a reliability of 99\% on the whole control signaling procedures, i.e., the reliability of transmitting all the four control messages depicted in Fig.~\ref{fig:controlled-frames}, assuming the information content detailed in~\cite{saggese2023impact}. Fig.~\ref{fig:reliability} shows the reliability as a heatmap function of the average \gls{snr} measured at the \gls{risc} and \gls{ue}, superimposing \gls{obcc} and \gls{ibcc} performance. Only the region satisfying the target reliability ($\ge 99$\%) is colored, where the minimum average \gls{snr} to achieve it is shown; the un-colored part of the heatmaps represents the \glspl{snr} values not satisfying the target reliability ($< 99$\%). 
In the case of \gls{ibcc}, \gls{oce} needs higher \glspl{snr} than \gls{bsw} to achieve the same reliability due to the higher information content of the control packets of the former. Specifically, the explicit transmission of each \gls{ris} element's response during the SET phase of \gls{oce} strongly impacts the minimum \gls{snr} required, implying that \gls{oce} is more prone to control errors. 
When \gls{obcc} is employed, the reliability of both approaches is enhanced as the control plane is leveraged for the transmission of critical control packets, i.e., codebooks and configurations. Therefore, \gls{obcc} is preferable to \gls{ibcc}, provided that its deployment is feasible in terms of resource availability and cost.

\section{Discussion \& Conclusions} \label{sec:conlusions}
We have presented design options for the control plane in \gls{ris}-equipped wireless systems. Compared to the number of theoretical and experimental studies focused on the \gls{phy} layer, the value of this work is that it elucidates a practical aspect of \glspl{ris}, essential from the viewpoint of system integration and standardization. The presented discussion provides the contours of the design space for \gls{ris} control plane. The analysis is necessarily simplified but still reveals the main trade-offs: on one side, the choice of \gls{ibcc} and \gls{obcc} mainly influences the reliability of the system while negligibly impacting the overhead; on the other, multiplexing-oriented communication approaches can attain better performance than diversity approaches, at the cost of being more susceptible to control errors and generating higher overhead. Control errors may be mitigated by investing in \gls{obcc} or by a robust resource allocation design for \gls{ibcc}. On the other hand, we remark that overhead may be reduced by exploiting prior information available to the decision-maker. For example, \gls{ue} localization information may help the \gls{bsw}~\cite{HRIS_DoA,Alexandropoulos2024:hris-magazine}, reducing the size of the BSW codebook and making it faster to converge to a desirable beam. The knowledge of the data plane statistics can help the \gls{oce}: in the presence of correlated or sparse data channel, the \gls{ce} codebook size can be reduced, decreasing, in turn, the \gls{csi} acquisition overhead. While not diving into more complicated aspects for simplicity of presentation, this article sets the basis for future work on more elaborate control plane designs that optimize various computation and communication parameters.

\bibliographystyle{IEEEtranNoURL}
\bibliography{IEEEabbr,bib}

\begin{IEEEbiography}
[{\includegraphics[width=1in,height=1.25in,clip,keepaspectratio]{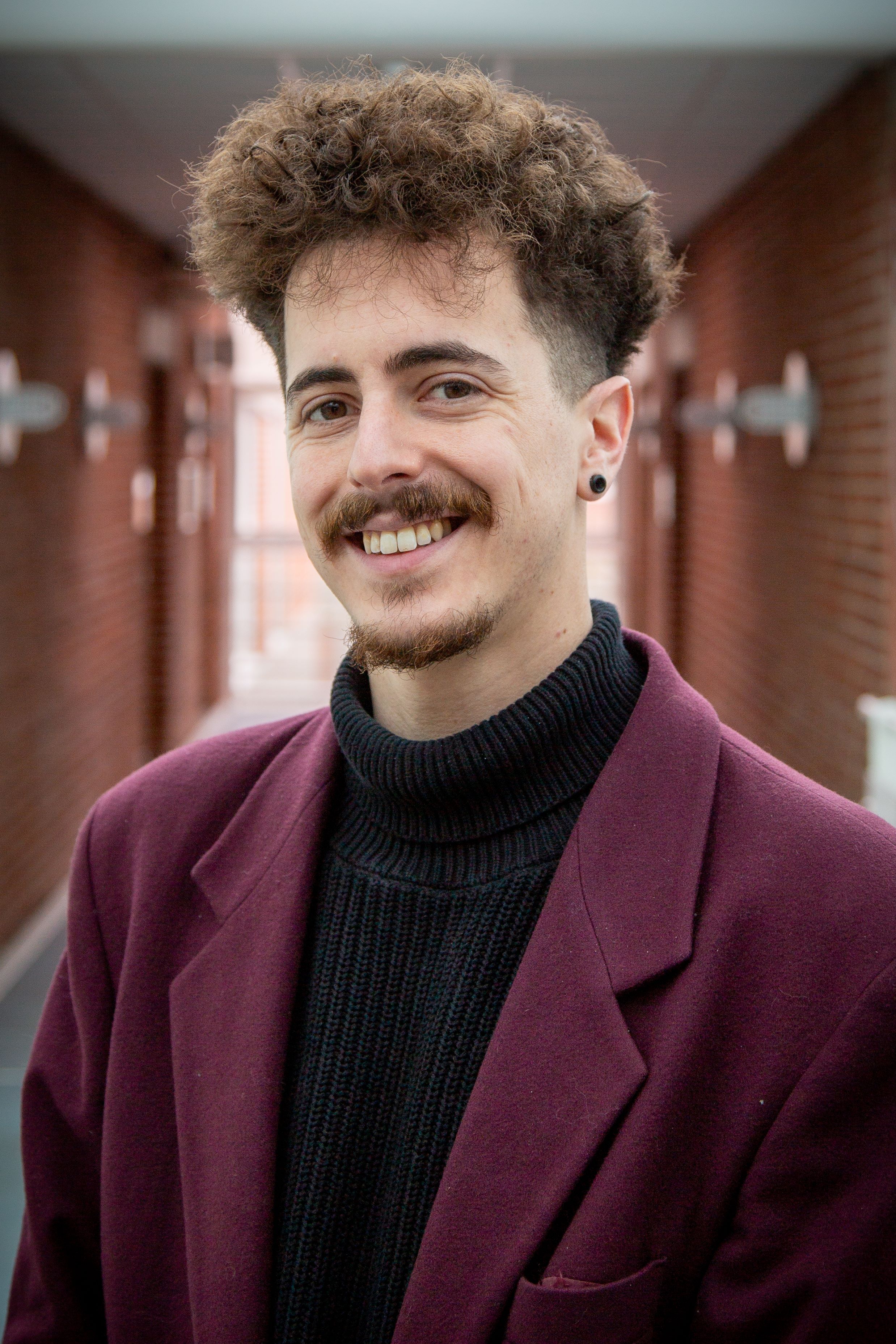}}]
{Fabio Saggese} 
(Member, IEEE) received the Telecommunication Engineering B.Sc. and the M.Sc. degrees, both summa cum laude, from the University of Pisa in 2015 and 2018, respectively. He received the Ph.D. title in Information Engineering from the University of Pisa in April 2022; the defended thesis was awarded by GTTI as the best PhD Thesis in Italy concerning Communications Technologies in September 2022. Since November 2021, he has worked at the Connectivity Section of the Electronic Systems Department, Aalborg University, Denmark. His research concerns resource allocation, MAC protocol design, and network slicing for future wireless networks.
\end{IEEEbiography}

\begin{IEEEbiography}[{\includegraphics[width=1in,height=1.25in,clip,keepaspectratio]{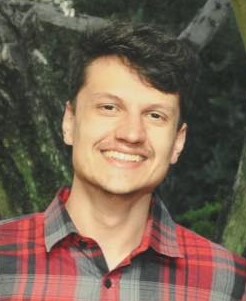}}]{Victor Croisfelt} (Graduate Student Member, IEEE) received his B.Sc. degree in Electrical Engineering from Universidade Estadual de Londrina, Londrina, Brazil, in 2018 and his M.Sc. degree in Electrical Engineering from Escola Politécnica of Universidade de São Paulo, São Paulo, Brazil, in 2021. He is currently pursuing his Ph.D. with the Connectivity Section of the Department of Electronic Systems at Aalborg University, Aalborg, Denmark. His current research interests are exploiting machine learning tools to design new and more efficient communication protocols.    
\end{IEEEbiography}

\begin{IEEEbiography} [{\includegraphics[width=1in,height=1.25in,clip,keepaspectratio]{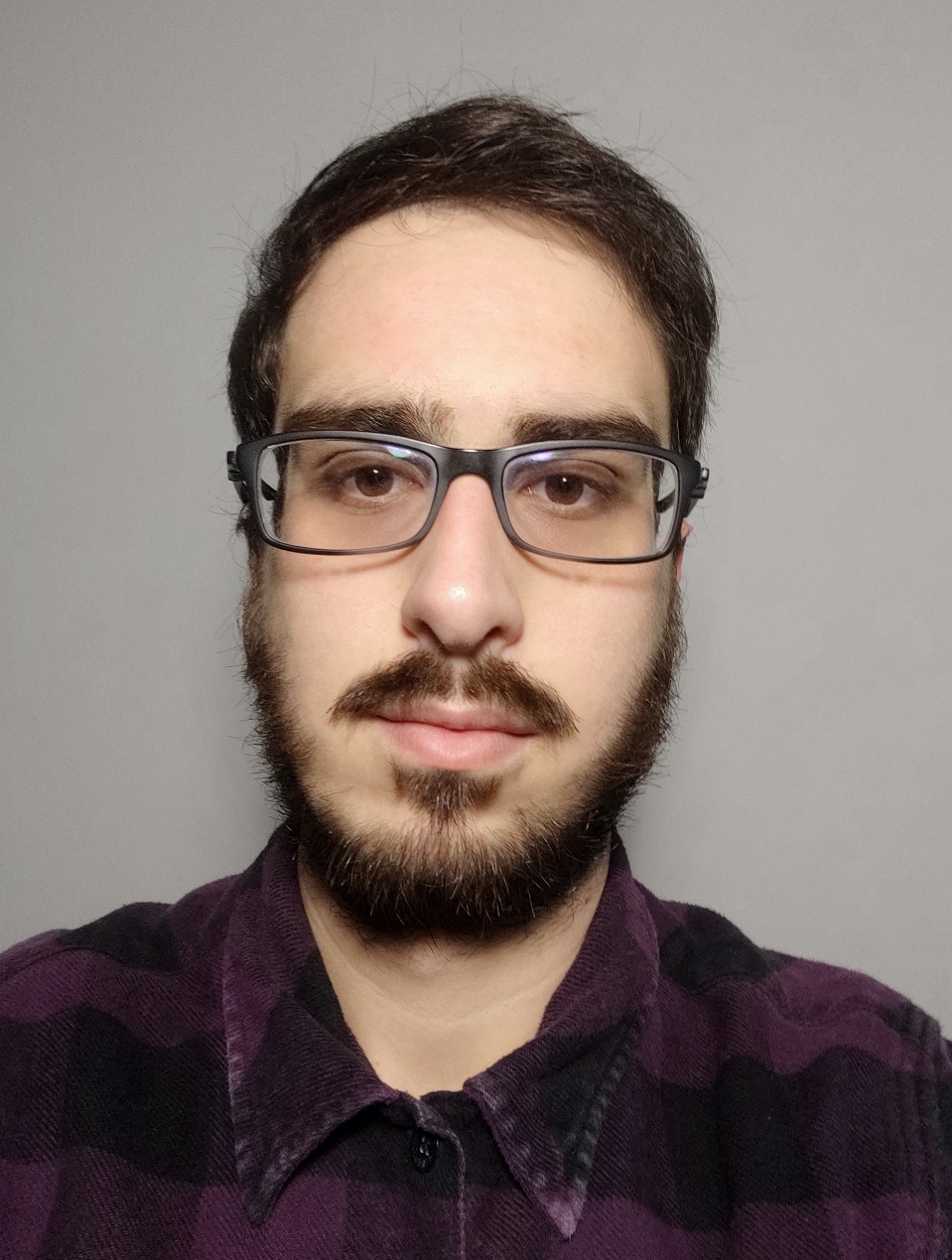}}]{
Kyriakos Stylianopoulos} (Graduate Student Member, IEEE) graduated from the Department of Informatics and Telecommunications, National and Kapodistrian University of Athens, Athens, Greece, in 2019. He received the M.Phil. degree from the University of Cambridge, Cambridge, U.K., in 2020. He is currently working toward the Ph.D. degree at the Department of Informatics and Telecommunications, National and Kapodistrian University of Athens, on the intersection of machine learning and beyond 5G wireless communications. His broader research interests include pattern recognition, deep learning, reinforcement learning, observational astronomy, reconfigurable intelligent surfaces and smart radio environments, and next-generation wireless communication networks.    
\end{IEEEbiography}

\begin{IEEEbiography}[{\includegraphics[width=1in,height=1.25in,clip,keepaspectratio]{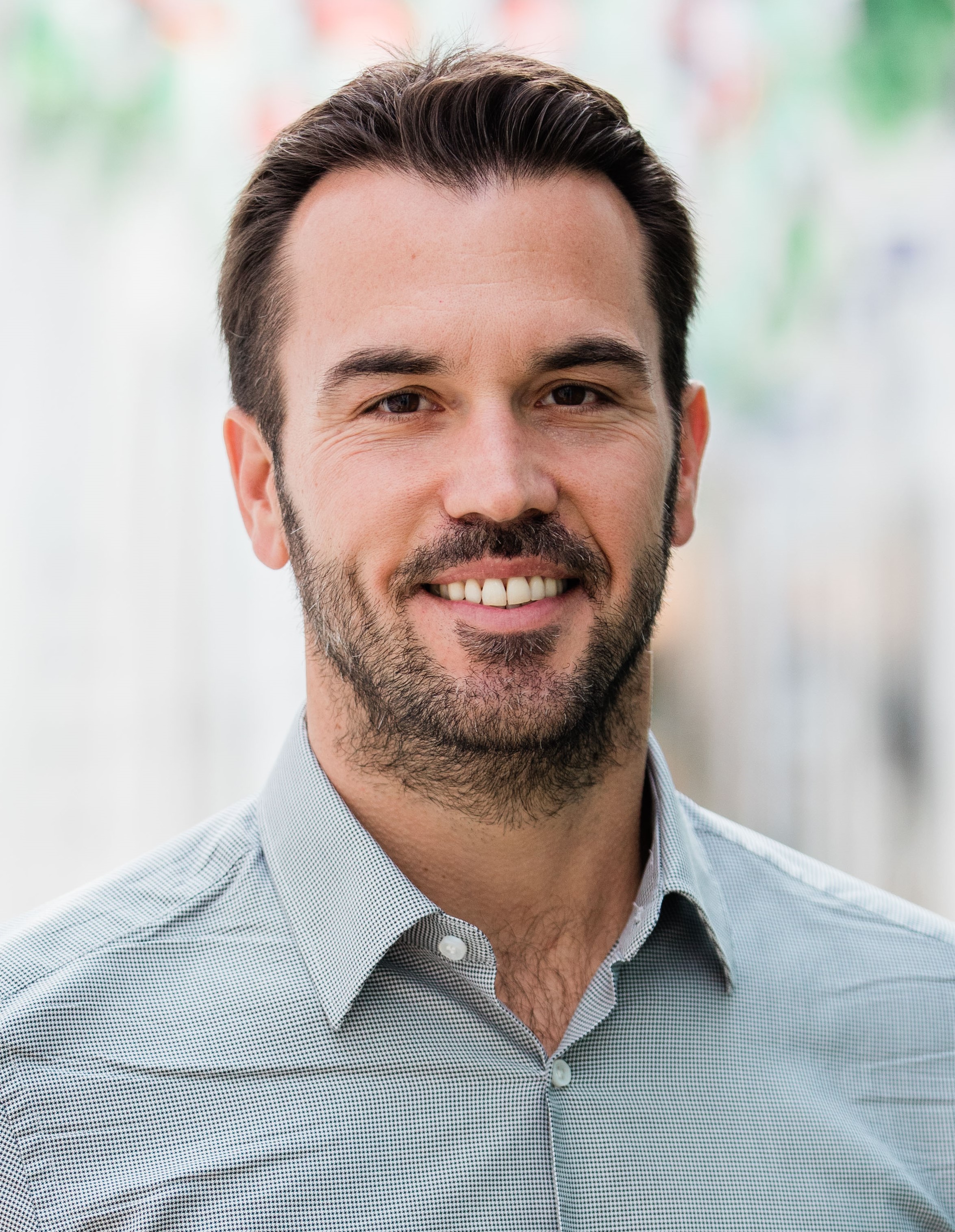}}]
{George C. Alexandropoulos} holds the Ph.D. degree from the University of Patras, Greece. He is currently an Associate Professor with the Department of Informatics and Telecommunications, National and Kapodistrian University of Athens, Greece, and an Adjunct Professor with the Department of Electrical and Computer Engineering, University of Illinois Chicago, Chicago, IL, USA. His research interests span the general areas of algorithmic design and performance analysis for wireless networks with emphasis on multi-antenna transceiver hardware architectures, full duplex radios, active and passive reconfigurable intelligent surfaces, integrated sensing and communications, millimeter wave and THz communications, as well as distributed machine learning algorithms.    
\end{IEEEbiography}

\begin{IEEEbiography}
[{\includegraphics[width=1in,height=1.25in,clip,keepaspectratio]{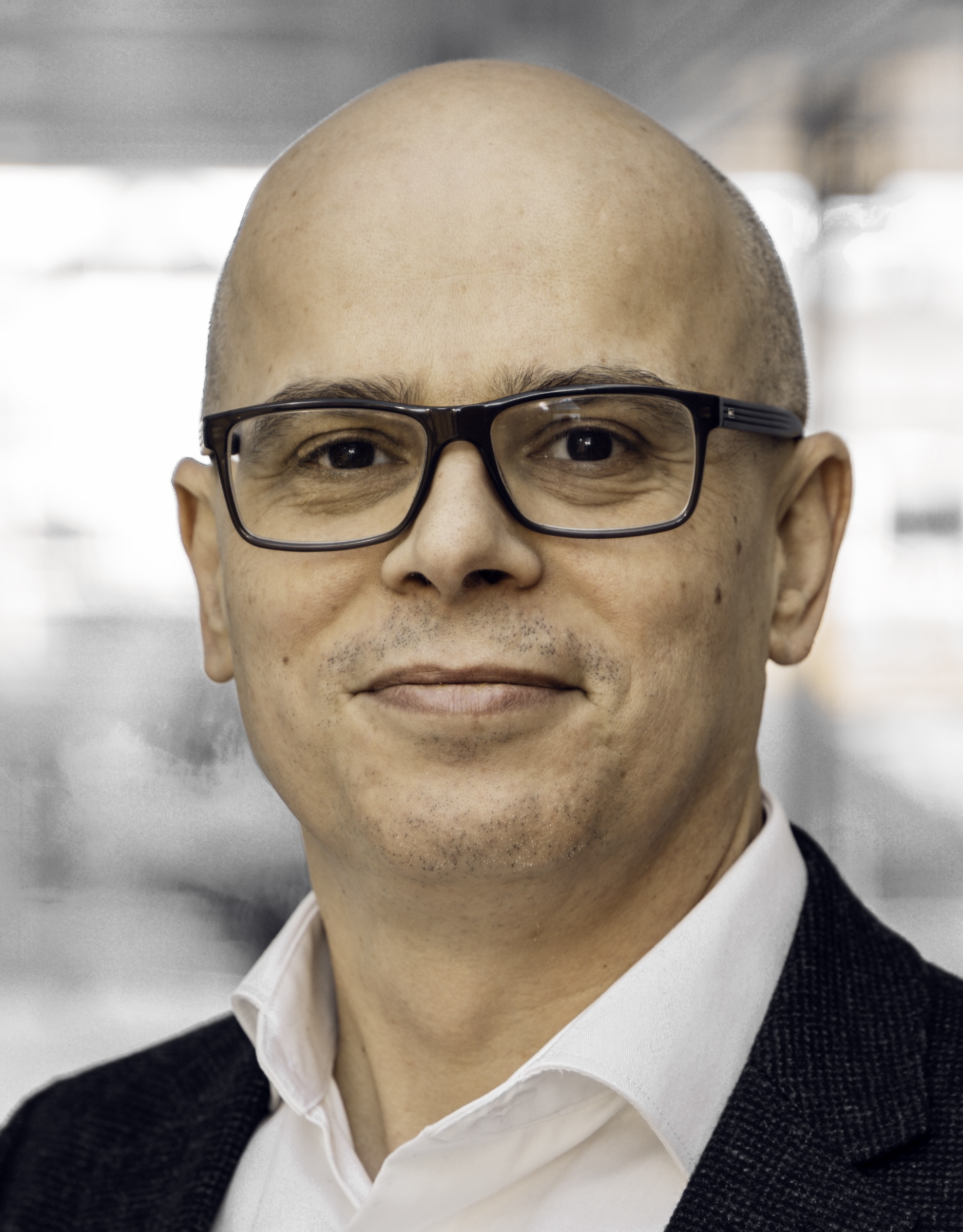}}]{Petar Popovski}
(Fellow, IEEE) is a Professor at Aalborg University, where he heads the section on Connectivity, and a Visiting Excellence Chair at the University of Bremen. He received his Dipl.-Ing and M. Sc. degrees in communication engineering from the University of Sts. Cyril and Methodius in Skopje and his Ph.D. from Aalborg University in 2005. He is currently an Editor-in-Chief of the IEEE Journal on Selected Areas In Communications and Chair of the IEEE Communication Theory Technical Committee. He authored the book “Wireless Connectivity: An Intuitive and Fundamental Guide.” His research interests are in the area of wireless communication and communication theory.
\end{IEEEbiography}

\end{document}